\documentclass[10pt]{mystyle}

\usepackage[linesnumbered,boxed]{algorithm2e}
\usepackage{wrapfig}
\usepackage{booktabs}
\usepackage{nicefrac}
\usepackage{lineno}
\title{Combining Compositional Data Sets Introduces Error in Covariance Network Reconstruction.}
\LAUR{LA-UR-23-32660}
\date{\today}

\author[1,2,*]{James D. Brunner}
\author[1]{Aaron J. Robinson}
\author[1] {Patrick S.G. Chain}
\affil[1]{Biosciences Division, Los Alamos National Laboratory, Los Alamos, NM, USA}
\affil[2]{Center for Nonlinear Studies, Los Alamos National Laboratory, Los Alamos, NM, USA}
\affil[*]{Email Address for Correspondence: jdbrunner@lanl.gov}
\affil[*]{Mailing Address for Correspondence:
James Brunner,
LANL MS-M888,
P.O. Box 1663,
Los Alamos, NM 87545
}

\begin{document}

\maketitle

{\bf Running Title:} Transkingdom Covariance Network Error.

\begin{abstract}
Microbial communities are diverse biological systems that include taxa from across multiple kingdoms of life. Notably, interactions between bacteria and fungi play a significant role in determining community structure. However, these statistical associations across kingdoms are more difficult to infer than intra-kingdom associations due to the nature of the data involved using standard network inference techniques. We quantify the challenges of cross-kingdom network inference from both a theoretical and practical viewpoint using synthetic and real-world microbiome data. We detail the theoretical issue presented by combining compositional data sets drawn from the same environment, e.g. 16S and ITS sequencing of a single set of samples, and survey common network inference techniques for their ability to handle this error. We then test these techniques for the accuracy and usefulness of their intra- and inter-kingdom associations by inferring networks from a set of simulated samples for which a ground-truth set of associations is known. We show that while two methods mitigate the error of cross-kingdom inference, there is little difference between techniques for key practical applications including identification of strong correlations and identification of possible keystone taxa (i.e. hub nodes in the network). Furthermore, we identify a signature of the error caused transkingdom network inference and demonstrate that it appears in networks constructed using real-world environmental microbiome data.
\end{abstract}


\section{Introduction}

Consider a scientist studying the flora and fauna of northern New Mexico. During their research, they count the plants and animals of the open plains of the valleys as well as the dense forests of the mountains, counting in discrete small sample areas. They report the abundances of rabbits and pi\~{n}on trees, with data shown in \cref{tab:rabbit_pinon_counts}. From these counts, we can see that pi\~{n}ons are spread approximately uniformly across the area, while rabbits prefer the open plains. From this, we may conclude that the amount of rabbits and the amount of pi\~{n}ons are not correlated, so there is no suggested relationship between rabbits and pi\~{n}ons. On the other hand, there exists a negative correlation between the amount of rabbits and the total amount of trees in an area.

\begin{table}[h]
    \centering
\begin{tabular}{ll|llll|llll}
\toprule
                  &  & \multicolumn{4}{l}{Plains} & \multicolumn{4}{l}{Forest} \\
Measure & Species &          &          &          &          &          &          &          &          \\
\midrule
Absolute & Rabbits &       62 &       59 &       70 &       54 &        2 &        5 &        4 &        3 \\
                  & Pi\~{n}ons &       10 &       11 &        9 &       12 &       11 &        8 &       13 &        7 \\
\midrule
Relative to  & Rabbits &     0.24 &     0.28 &     0.32 &     0.23 &     0.01 &     0.02 &     0.01 &     0.03 \\
      Kingdom            & Pi\~{n}ons &     0.82 &     0.91 &     0.78 &     0.83 &     0.09 &     0.11 &     0.08 &     0.12 \\
\midrule
Relative to  & Rabbits &     0.23 &     0.26 &     0.30 &     0.22 &     0.01 &     0.02 &     0.01 &     0.02 \\
    Total              & Pi\~{n}ons &     0.04 &     0.05 &     0.04 &     0.05 &     0.03 &     0.02 &     0.02 &     0.04 \\
\bottomrule
\end{tabular}
    \caption{In the absolute counts of rabbits and pi\~{n}ons, clearly, there is no actual correlation between the two organisms. However, relative abundance with respect to the taxa's kingdom (i.e. animal or plant) make it appear that rabbits and pi\~{n}ons are positively correlated. When considering the relative abundance relative to total biomass, we see that this effect was an artifact of mishandling compositional data.}
    \label{tab:rabbit_pinon_counts}
\end{table}

Now, suppose that the data is presented compositionally, so that we see only the relative abundance (among animals) of rabbits and the relative abundance (among trees) of pi\~{n}ons. The negative correlation between rabbits and total tree population appears instead as a positive correlation between rabbits and pi\~{n}on. However, if the data is presented relative to total organisms (effectively accounting for the difference in total plants and total animals across samples) then the correlation between rabbits and pi\~{n}ons again disappears.

This relatively simple example demonstrates a profound problem for scientists studying the microbiome, where instead of animals and trees we are concerned with, for example, bacteria and fungi. In this setting, taxonomic information is compositional by nature\cite{tsilimigras2016compositional,gloor2017microbiome}, and there exists only limited ways to compute absolute biomass of taxa\cite{boshuizen2023pitfalls,swift2023review}. Furthermore, data on two or more kingdoms of taxa (called ``transkingdom" data) are often collected with separate methods for each kingdom. The most common example of this is the use of 16S rRNA amplicon sequencing to identify the bacteria in a sample paired with ITS rRNA amplicon sequencing to identify the fungi. The result is the relative abundance of each bacterial strain identified among bacteria, and separately the relative abundance of each fungal strain identified among fungi. Notably, the relative abundance of each taxa among the complete set of taxa is unknown. While established techniques can be used to handle compositional data\cite{aitchison1982statistical,aitchison1992criteria,aitchison2000logratio}, these techniques are designed to work when the composition is known relative to the total data set. When dealing with transkingdom data, we must therefore take into careful consideration if we are properly handling the compositional data.

\begin{figure}
    \centering
   \includegraphics[scale = 0.8]{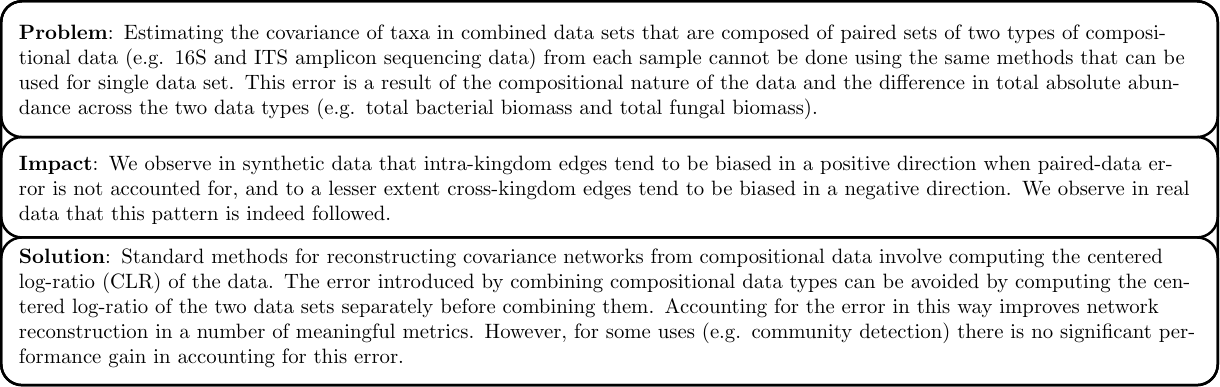}
    \caption{Short summary of our results.}
    \label{fig:textbox}
\end{figure}

Despite the difficulties, understanding microbiomes will require an understanding of the interactions that occur across kingdoms \cite{bergelson2019characterizing,lee2022cross}. In an effort to understand these interactions, many researchers have used paired sets of 16S and ITS data to infer interaction networks that include both bacteria and fungi. These networks have been used to study how the interactions between bacteria and fungi impact their plant hosts to promote plant growth\cite{yuan2021fungal}, to identify possible keystone taxa across kingdoms\cite{agler2016microbial,banerjee2016network}, and to demonstrate a fungal role in human disease associated dysbiosis\cite{sokol2017fungal,lemoinne2020fungi}.  It is therefore essential that the error introduced by combining compositional data types be well understood. \Cref{fig:textbox} summarizes the problem and our results.

\section{Materials and Methods}

\subsection{Computing the error introduced by combining paired compositional data sets}

In the scenario of the rabbits and pi\~{n}ons, there are two compositional data sets, one which measures animals and one which measures plants. These sets are ``paired" with each other in the sense that each vector in one data set, representing for example the number of each animal species counted in an area, can be paired with a vector in the other set representing the number of each plant species counted in the same area. These two vectors, coming from the same discrete sample area, can be thought of as a pair in the combined paired data set.

In the study of microbiomes, a scientist may have counts of bacterial taxa (called ``operational taxonomic units" or ``OTUs") as determined by 16S amplicon sequencing from a set of, e.g., soil samples as well as the counts of fungal OTUs as determined by ITS amplicon sequencing of the same samples. Each soil sample provides two data vectors which we may consider as a pair. However, it is important to note that each member of the pair is compositional, meaning that the counts only have meaning relative to the total count of the vector. In other words, the true abundance of each OTU is not the reported count, but is instead proportional to the reported count. This means that if the relative abundance vector from the 16S counts from sample $i$ is $\b{x}^i = (x^{i}_1,x^{i}_2,...,x^i_n)$ then the true abundances are 
\[
\b{X}^i = (X^i_1,...,X^i_n) \approx (q^i x^i_1,...,q^ix^i_n) = q^i \b{x}^i
\]
for some unknown $q^i$ and likewise if $\b{y}^i$ is the vector of relative ITS counts, then the true fungal abundances are 
\[
\b{Y}^i \approx r^i \b{y}^i
\]
for some unknown $r^i$ (and these equations become exact in the limit that total read count approaches infinity). Furthermore, the $q^i$ and $r^i$ values are not constant across the two data sets, but instead are different for each sample. The established techniques for dealing with this compositional nature for, for example, the 16S data set involve log-ratio transformations that are consistent regardless of the value of $q^i$ \cite{aitchison1982statistical,aitchison1992criteria,aitchison2000logratio}, but depend on the fact that $q^i\b{x}^i$ is a scalar multiple of $\b{X}^i$. This means that these techniques fail when implemented on the concatenated vector $(\b{x}^i,\b{y}^i)$ unless $q^i=r^i$ (or the ratio $\frac{q^i}{r^i + q^i}$ is known). For example, one popular technique for computing the correlation between taxa $a$ and taxa $b$ across a data set relies on computing 
\begin{equation}
    \text{Var}\left[\log\left(\frac{X_a}{X_b}\right)\right] = \text{Var}\left[\log\left(\frac{q x_a}{q x_b}\right)\right] = \text{Var}\left[\log\left(\frac{x_a}{x_b}\right)\right]
\end{equation}
which is the same for the absolute data as it is for the relative. However, if we attempt this with two taxa in different data sets (e.g. a bacteria and fungi), we see that 
\begin{equation}\label{eq:xking_err}
    \text{Var}\left[\log\left(\frac{X_a}{Y_b}\right)\right] = \text{Var}\left[\log\left(\frac{q x_a}{r y_b}\right)\right] \\= \text{Var}\left[\log\left(\frac{x_a}{x_b}\right)\right] + \text{Var}\left[\log\left(\frac{q}{r}\right)\right] + 2\text{Covar}\left[\log\left(\frac{q}{r}\right),\log\left(\frac{x_a}{y_b}\right)\right].
\end{equation}
The cancellation that the method relies on does not occur, and error is introduced from the differences in total fungal vs. bacterial biomass in each sample.

Using the centered-log-ratio transform on the data directly causes similar issues, and the error can again be computed algebraically as a function of the ratio of the scaling factors $q^i$ and $r^i$. To correct for these errors directly, it is necessary to estimate the ratio of the total absolute abundances between the two data sets (e.g. total fungi to total bacteria), meaning $\frac{q^i}{r^i}$, for each pair of samples. 

Probabilistic methods offer a possible alternative to estimating total biomass\cite{tipton2018fungi,aktukmak2022graphical}. In Tipton et al.\cite{tipton2018fungi}, the authors make the observation that if the two data sets are transformed according the the centered-log-ratio separately, they can then be concatenated and a covariance matrix can be computed, with the result being a good approximation to the covariance matrix of the actual absolute log-abundances. This means that any method which relies on estimating the covariance of the log-abundance data, such as the Guassian LASSO method \cite{kurtz2015sparse}, can be adapted for use with paired data sets.

\subsection{Generating Synthetic Data}

To test the above theoretical problems involved in network reconstruction with paired data sets, we developed a simple algorithm for constructing synthetic data from a known ``ground-truth" covariance matrix. To do this, we start with a random Power-law graph and construct a positive definite covariance matrix that matches the sparsity pattern of the graph but has positive and negative non-integer entries. This construction ensures that we have a sparse, small-world network of positively and negatively correlated nodes, representing synthetic taxa. The simulated exact absolute abundance of each taxa in each sample is then generated by drawing samples from a log-normal distribution using the ground truth covariance matrix and mean log-values drawn uniformly from the interval $(-4,4)$. We then generate synthetic data by simulating sequencing experiments. To do this, we draw $R_i$ ``reads" from each sample, with $R_i$ some hyper-parameter chosen from a normal distribution with mean $100000$ and standard deviation $10000$ for each sample. Each simulated read is generated by drawing from the set of taxa with a discrete probability distribution equal to the relative abundance of the taxa in the exact sample. This simulates the real read process in which each amplicon that is counted has a probability of being classified as from some particular taxa approximately equal to the relative abundance of that taxa. 

By generating the data from a log-normal distribution, we construct a synthetic data set from which it should be relatively easier to reconstruct a network in comparison to real data. Real data, unlike our simulated data, may be confounded by many factors, including environmental and climate effects, variable copy number of amplified genes (e.g. the 16S rRNA gene), or even a lack of a true underlying network of interactions. Furthermore, all of the simulated samples are drawn from the same underlying distribution, which may be thought of as taking samples from identical environments and ignoring spatial heterogeneity. This scenario therefore represents a near ``best-case scenario" for network reconstruction. The results that follow should therefore be taken as ``necessary but not sufficient" in judging a particular method to be useful in network reconstruction. It may be interesting to repeat the experiments with additional methods for generating synthetic data which simulate some of the real-world confounding variables that our method ignores.

To simulate paired compositional data sets, we randomly split the taxa into two groups representing two kingdoms of life before drawing synthetic reads. Then, for each exact sample, we generate two independent sets of reads as described above --- one for each of the two kingdoms. This means that for each sample, we generate a pair of synthetic data representing the relative abundances of each of within its group. This simulates the scenario in which the scientist counts animals and plants (or, in the case of microbiome analysis, bacteria and fungi) separately and has no estimate of the total biomass of either group. 

\subsection{Comparison Using Real Data}

We used a paired ITS \& 16S compositional data-set published by de Vries et al.\cite{de2018soil} to reconstruct co-occurrence networks and inspect the results for the characteristics we identified during our experiments with synthetic data. In order to save computational time, we used only the top $2\%$ most abundant fungal taxa (164 taxa) and top $1\%$ most abundant bacterial taxa (180 taxa). Different proportions were used simply so that we could use a similar number of taxa for each kingdom. We subdivided the data set according to the experimental groups presented in the original publication. In that experiment, different mesocosms were planted with different plant community compositions, and two possible drought conditions (drought or no drought) present in the data. In order to reduce the effect of differential interactions, we constructed networks for a subset of the available samples which had \emph{Anthoxanthum odoratum} as the dominant plant and did undergo drought.

\subsection{Comparing Reconstruction Methods}

There are many techniques available for reconstructing the network of interactions between microorganisms in the microbiome. Some of these involve fitting the parameters of a dynamical model, e.g. the generalized Lotka-Volterra model \cite{bucci2016mdsine}, while others make use of functional information derived from genomic data \cite{diener2020micom,kim2022resource}. These functional and dynamic methods rely on dense time-course data or often incomplete and erroneous gene-finding and annotation, and so are expensive to employ. For this reason, the most commonly used methods involve statistical relationships between taxa in a set of samples.

Statistical network reconstruction is appealing because it requires only abundance data across a set of samples, with no requirement for time-longitudinal sampling or detailed genomic analysis such as gene annotation. However, microbiome data is compositional, rather than absolute, so computing correlations directly will lead to many spurious connections in a network\cite{aitchison1982statistical,aitchison1992criteria,gloor2017microbiome,tsilimigras2016compositional}. The problem of reconstructing a network from compositional data has led to the creation of a set of techniques designed to eliminate these spurious edges\cite{weiss2016correlation}. These techniques are generally based on the idea of log-ratio transformations\cite{aitchison1982statistical}, which produce the logarithm of a sample that has been normalized by some function of that sample (e.g. the geometric mean). Some of these, including the tool ``SparCC"\cite{friedman2012inferring}, estimate correlations directly using the log-ratio of read counts. Others, such as the Guassian LASSO method (which is implemented in the popular ``Spiec-Easi" tool\cite{kurtz2015sparse}) and other probabilistic models\cite{fang2015cclasso,biswas2016learning}, use parameter fitting procedures to infer the parameters of an underlying distribution by maximizing the likelihood of log-ratio transformed data. Each of these methods was developed to avoid the bias introduced by the compositional nature of the data for a single data set.

To investigate the reconstruction of networks from paired data sets (e.g. 16S and ITS data), we used six methods for inferring covariance between taxa. These were the following: 
\begin{enumerate}
    \item simply computing the empirical covariance of the log-values of the simulated counts, which we call the ``Log-Covariance" method,
    \item computing the covariance the centered-log-ratio transform of the simulated counts after concatenating the two data sets, which we call ``CLR-Mixed",
    \item computing the covariance of the centered-log-ratio transform of the simulated counts with the transform taken separately on the two data sets, which we call ``CLR-Split",
    \item using the SparCC \cite{friedman2012inferring} method after concatenating the two data sets,
    \item using the Guassian LASSO (GLASSO) as introduced in the R package ``Spiec-Easi" \cite{kurtz2015sparse} method after concatenating the two data sets, which we call ``GLASSO-Mixed",
    \item and using the GLASSO method but fitting to the ``split" centered-log-ratio transform of the simulated data, as in \cite{tipton2018fungi}, which we call ``GLASSO-Split".
\end{enumerate}

Note that the first three of these methods simply compute correlations directly from the count data, and only differ in the application of the centered log-ratio. These represent the baseline ``preprocessing" that is necessary to avoid the errors introduced by compositional data and paired data sets. Differences in results between these three methods demonstrate the importance of the CLR transformation and the importance of carrying out this transformation correctly for paired data sets (i.e. separately). The last three methods represent the two popular categories of more sophisticated network reconstruction --- constrained inference of covariance (SparCC) and fitting to an assumed distribution (GLASSO). To our knowledge, SparCC\cite{friedman2008sparse} and GLASSO (as implemented in ``Spiec-Easi")\cite{kurtz2015sparse} are the most widely used network reconstruction methods for microbiome data.

Unlike the other methods, the GLASSO method attempts to infer the inverse covariance matrix of the underlying distribution. In order to compare ``apples to apples", we took the matrix inverse of the GLASSO fitted result so that all six methods gave a covariance matrix. In our experiments, we used $500$ total simulated taxa and $75$, $150$, or $300$ simulated samples. For each number of samples, we separately created $15$ synthetic data sets and fit a network to each with each of the six network inference methods. We quantified intra- and cross-kingdom edge bias using the mean edge strength relative to the strongest edge of the intra- and cross- kingdom edges.

Noting that absolute root-mean-square or total error is skewed by the GLASSO methods lower average covariance magnitude, we quantify the absolute edge error by using the coefficient of determination (often called the ``$R^2$ value") from linear regression between true and fit edges. Moving beyond the simple error of all edges, we evaluated how well these network algorithms could accomplish two tasks. These are (1) identifying edges, especially identifying strong edges which may indicate meaningful interaction, and (2) identifying hub nodes in the network, which are often assumed to correspond keystone taxa in a population. We investigated both of these questions by treating the network as a classifier for edge detection and hub detection, respectively, and used common metrics to study classifier performance.

\section{Results}

\begin{figure}
    \centering
    \includegraphics[scale=0.175]{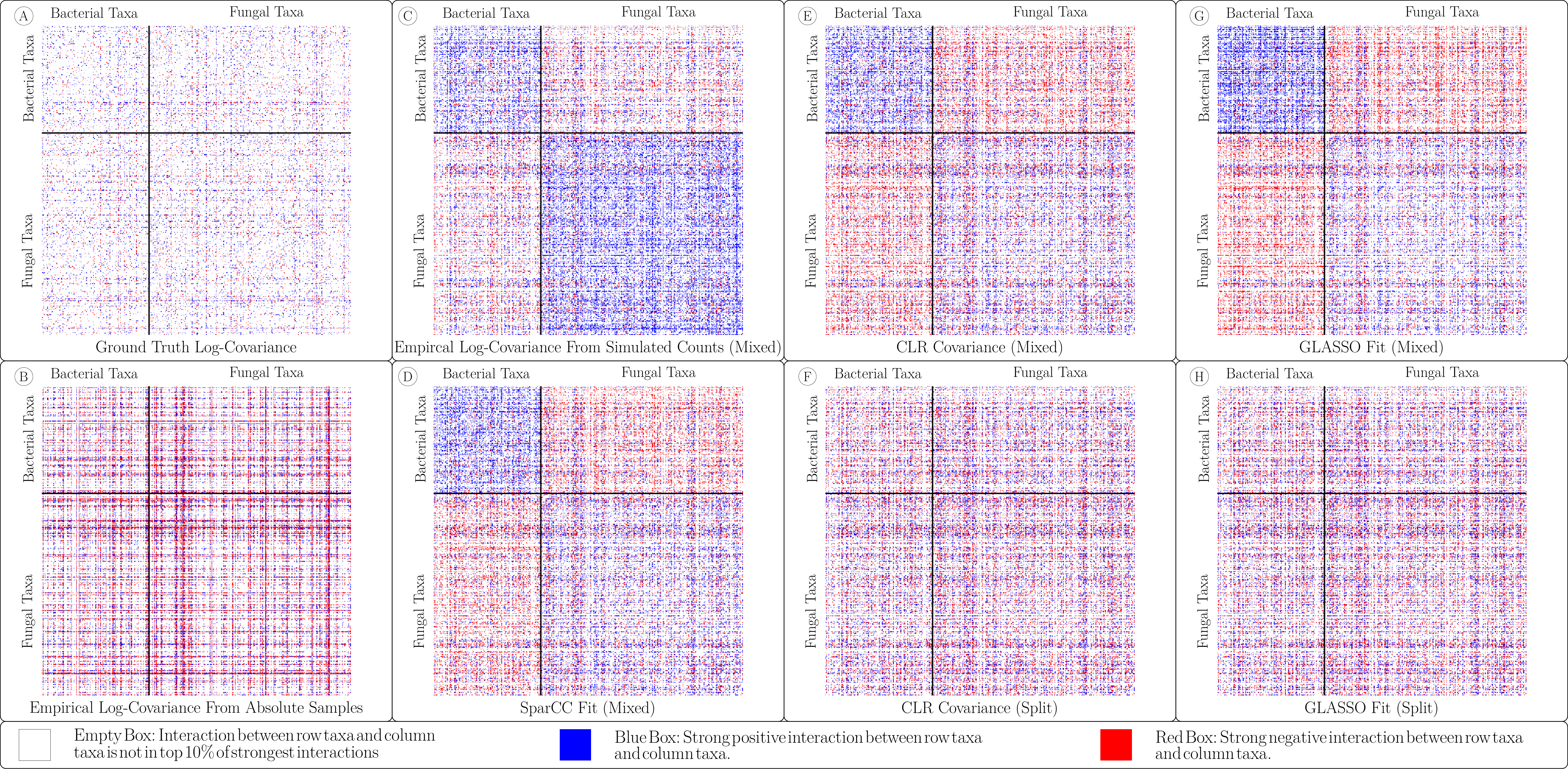}
    \caption{The ground truth underlying covariance matrix (A) used to generate the synthetic data along with the empirically computed log-covariance matrix of the absolute abundance data (B) (150 samples) and networks generated using the six methods for reconstructing the covariance matrix. The top $10\%$ of edges by absolute strength are shown color coded as red for a negative interaction and blue for a positive interaction. Networks (A) and (B) cannot be computed with real data. Networks (C) and (D) were constructed with methods that cannot be corrected for transkingdom data. Networks (E) and (F) are both constructed in using CLR-covariance, but in network (F) the CLR transform was computed separately for each of the two data sets in the pair. Likewise, networks (G) and (H) are constructed in a similar manner, using the GLASSO method, but only network (H) corrects for the error introduced by using paired data. Black bars separate the taxa from the two kingdoms, so that the top left and bottom right square blocks represent intra-kingdom edges, while the top right and bottom left rectangles represent cross-kingdom edges.}
    \label{fig:nets}
\end{figure}

Of the six methods, four (Log-Covariance, CLR-Mixed, SparCC, and GLASSO-Mixed) do not address the problems that arise when using paired data sets. Indeed, in \cref{fig:nets}, we observe an obvious artifact of concatenating the paired data in each of these networks. That is a strong set of positive correlations within each individual data set and a strong set of negative correlations across the two sets. These correlations do not arise due to actual relationships between the taxa, but instead arise the ratio of total biomass of two groups. 

\begin{figure}
    \centering
    \includegraphics[scale=0.35]{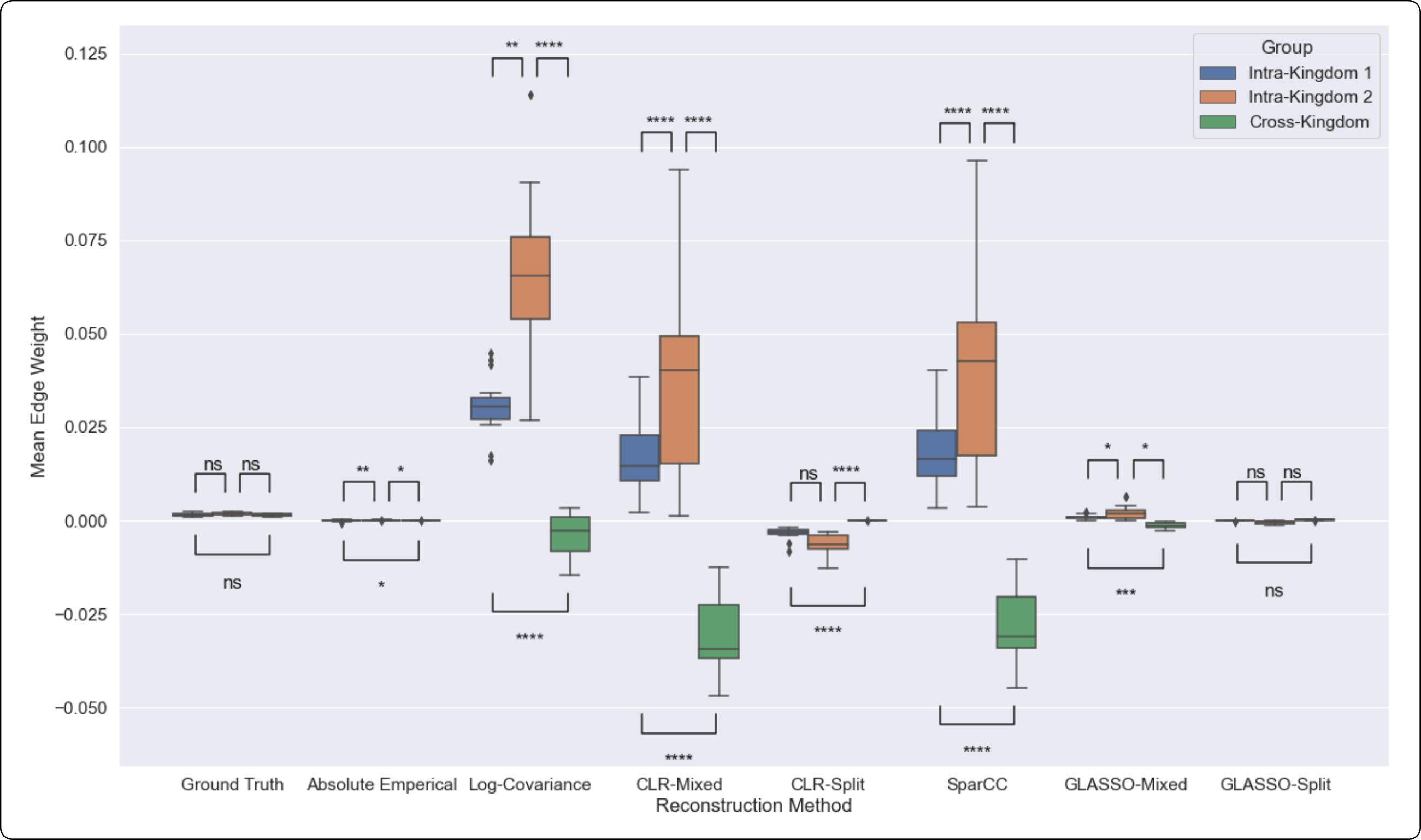}
    \caption{To quantify the bias apparent in \cref{fig:nets}, we computed the mean edge strength for intra- and cross-kingdom edges. Methods that do not address the error of combining data sets show a positive covariance bias within a data set and slight negative bias in cross-kingdom edges. Each of the 15 network reconstruction experiment used 150 samples from a randomly generated ground truth covariance to generate a network using each of the 6 methods. Note that the significance indicated is the average across all trials of the significance of the difference between edge strengths across groups of edges.}
    \label{fig:meanedge}
\end{figure}

We observe this bias across the 15 experiments, as shown in \cref{fig:meanedge}. The sparsity of the GLASSO method obscures this bias compared the other methods tested (especially the log-covariance and CLR methods) in the GLASSO-Mixed networks, but otherwise the trend is consistent. The mean intra-kingdom edge is positive while the mean cross-kingdom edge is negative in the log-covariance, SparCC, CLR-Mixed, and GLASSO-Mixed networks, but not in the true covariance matrices, the covariance of absolute data, or the methods that account for combining data sets.

\subsection{Edge Error}

\begin{figure}
    \centering
    \includegraphics[scale=0.23]{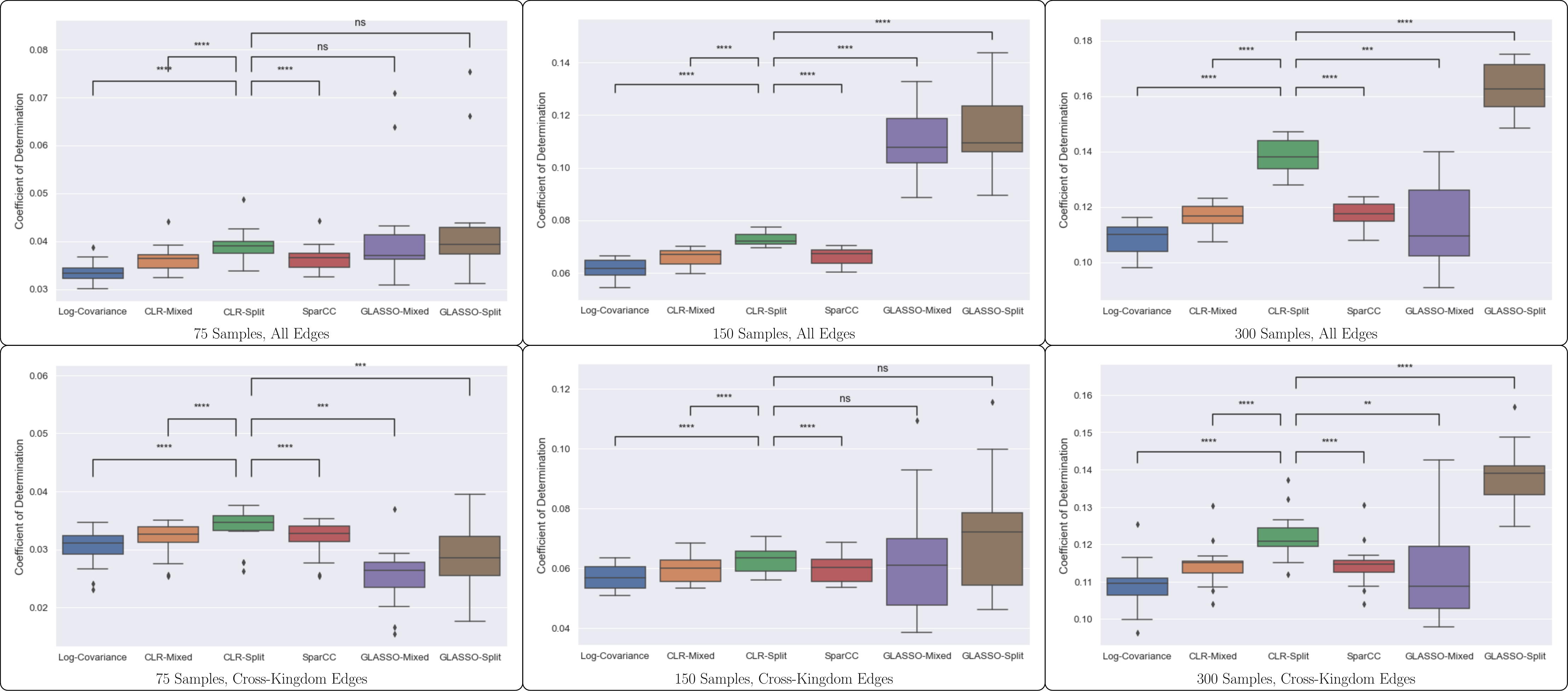}
    \caption{While methods that account for the error introduced by pairing data types (CLR-Split and GLASSO-split) have improved performance in fitting edges in a network, as evidenced by higher coefficients of determination in linear regression, this effect is not extremely pronounced, especially with fewer samples. When including only cross-kingdom edges, the difference between methods is less pronounced. We present the significance of the difference between the ``CLR-Split" method and the other methods to emphasize the importance of addressing the error of cross-kingdom covariance estimation.}
    \label{fig:cdet}
\end{figure}

While visual inspection of the inferred adjacency matrices clearly shows the bias introduced by neglecting to account for the two separate data types, quantitative analysis reveals that these effects are not extremely strong. We observe in \cref{fig:cdet} that the GLASSO-Split and CLR-Split methods, the two methods that correct for paired-data, do perform better than the others in the coefficient of determination, with the GLASSO-Split method performing best. The scale of this improvement does depend on the number of samples used.

\subsection{Adjusting the Ground-Truth Network Structure}

There is no \emph{a priori} reason to assume that cross-kingdom interactions occur at the same density as intra-kingdom edges, nor that they are just as likely to be positive as negative. We investigated the edge error of network reconstructions from data simulated with ground-truth covariance that had more or less dense cross-kingdom covariance, as well as positively and negatively biased cross-kingdom edges. The pattern of accuracy between the six network reconstruction methods was mostly unchanged by changing the sparsity pattern for more or less sparse cross-kingdom covariance. The only observable difference was that the ``GLASSO" methods, which performed significantly better than other methods with an even sparsity pattern, did not perform significantly better than the ``CLR-Split" when cross-kingdom covariance was sparse.

\begin{figure}
    \centering
    \includegraphics[scale=0.3]{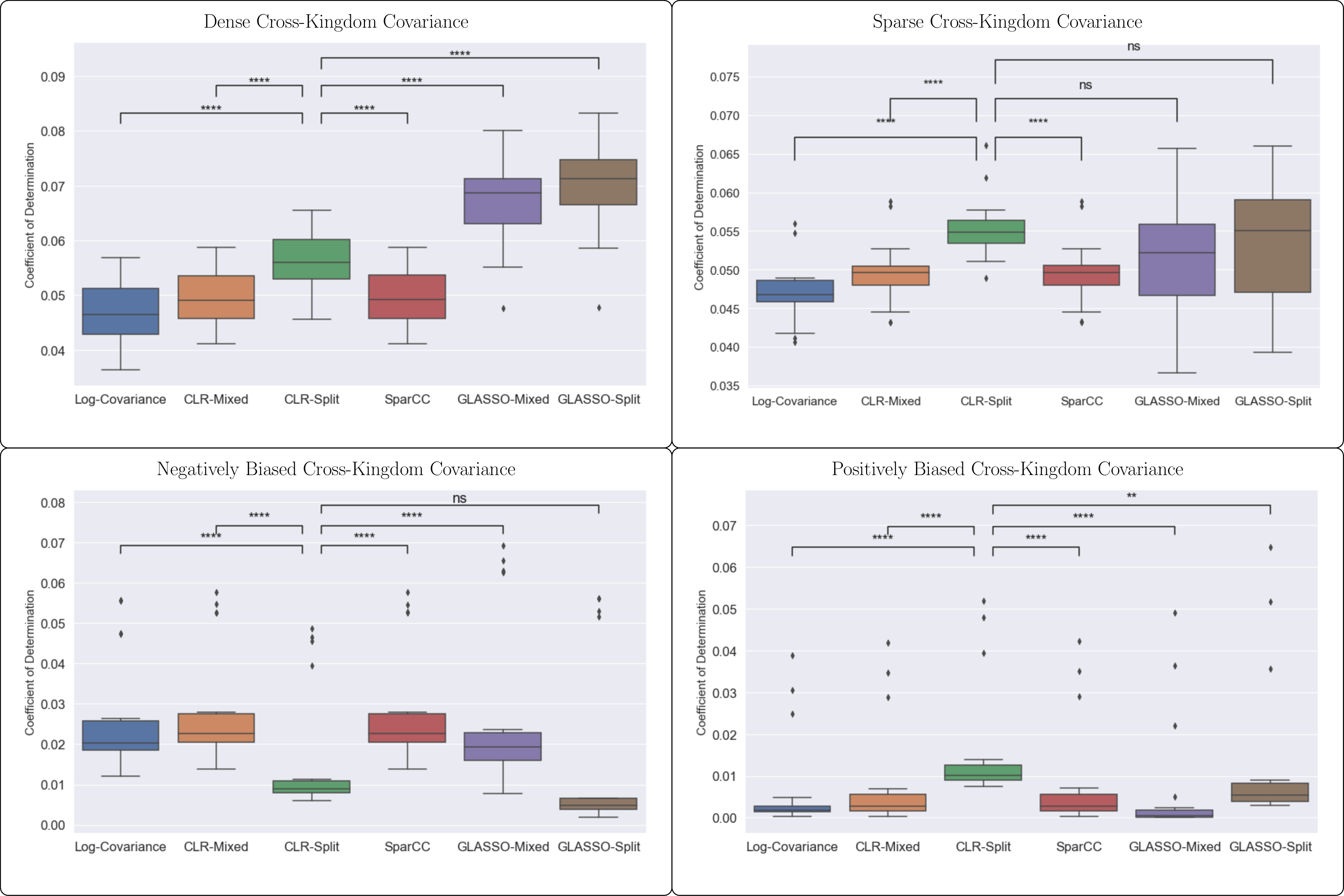}
    \caption{Adjusting the structure of the underlying ``ground truth" so that the cross-kingdom covariance was either denser or sparser than the intra-kingdom covariance had little relative effect on the methods, although all methods performed more poorly. However, underlying ``ground truth" experiments with edge bias did show different results. When cross-kingdom edges were more likely to be negative, the methods that account for the error introduced by pairing data types (CLR-Split and GLASSO-split) actually fair worse than other methods, which introduce an artificial bias that happens to match the correct bias. When cross-kingdom edges were more likely to be positive, methods that did not account for paired-data error performed very poorly.}
    \label{fig:metaparms}
\end{figure}

Introducing bias in the cross-kingdom covariance, on the other hand, did have an interesting result. From \cref{fig:nets} and \cref{fig:meanedge}, we can observe that the error introduced by combining paired data sets tends to appear as a negative bias in cross-kingdom edges and positive bias in intra-kingdom edges. Stronger performance of those methods that do not account for this error is therefore to be expected when the underlying ground-truth shares this bias. Indeed, \cref{fig:metaparms} shows that these methods perform more strongly relative to the methods that account for the error. More surprisingly, the methods that account for the error introduced by combining paired data sets perform significantly worse than those that do not. These two methods, ``CLR-Split" and ``GLASSO-Split" perform similarly when the bias is reversed, while the other methods perform very poorly in that case.

\subsection{Edge Detection and Accuracy}

\begin{figure}
    \centering
    \includegraphics[scale=0.25]{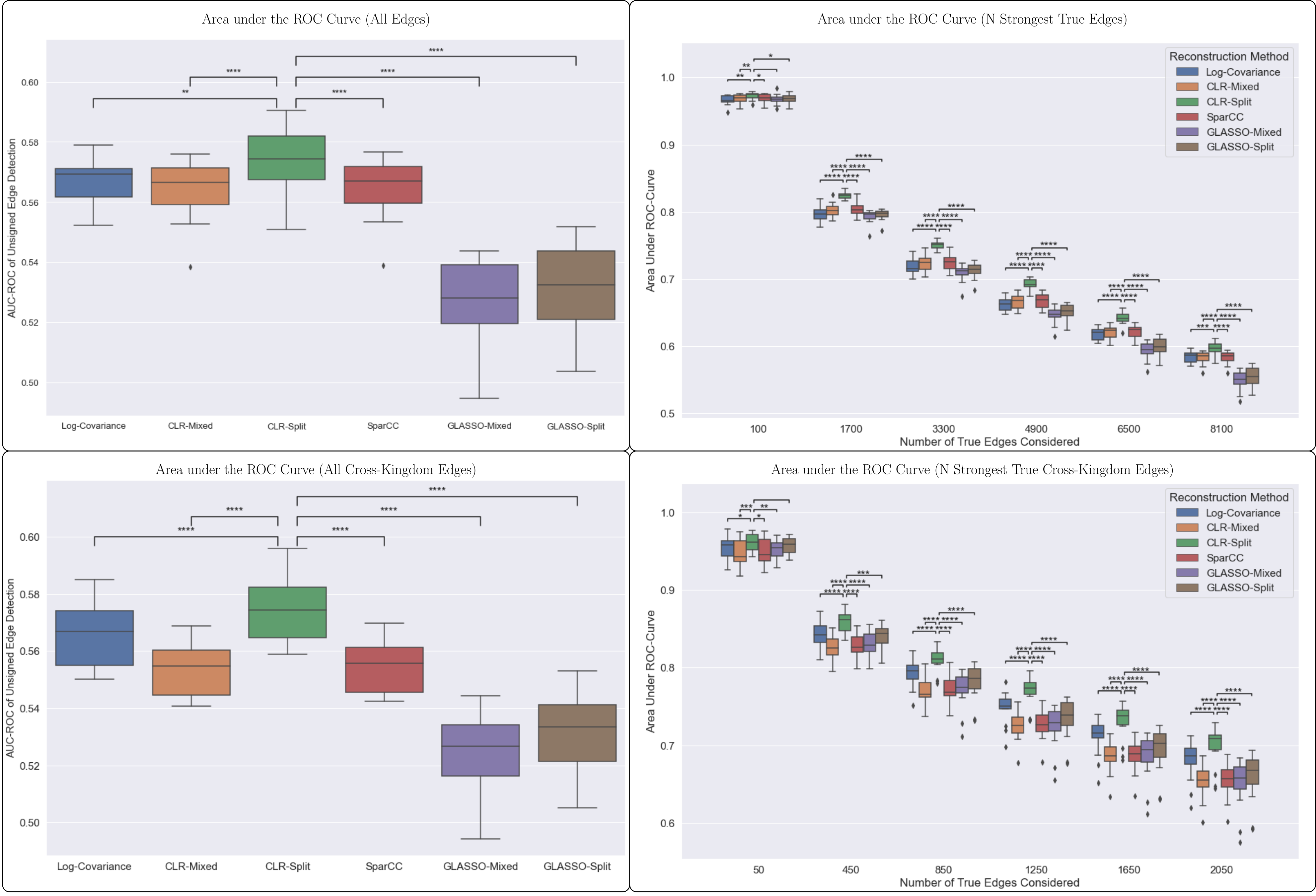}
    \caption{Area under the ROC curves reveal that simple CLR works well as a classifier for experiments with $150$ samples after accounting for combined data sets (left). If only the strongest true edges are considered to be ``true", then both of the methods that account for combined data sets outperform the other methods. We present the significance of the difference between the ``CLR-Split" method and the other methods to emphasize the importance of addressing the error of cross-kingdom covariance estimation.}
    \label{fig:roc}
\end{figure}

One major use of network inference is to detect possible interactions between taxa. For this reason, we assessed each method's ability to detect edges using two metrics. The first is metric treats the network as a classifier for each possible edge between two taxa. This means that if the fit score for a possible edge is above some threshold in magnitude, then the edge is classified as present. Note that this scoring ignores the sign of the edges. We then compute the receiver operating characteristic (ROC) for the classifier, and finally score the network methods using the area under the ROC curve. As \cref{fig:roc} shows, the CLR-Split method performs significantly better than the other methods for all edges as well as only cross-kingdom edges. Interestingly, GLASSO-Split method does not work well as a classifier unless we restrict the edges considered to be ``true" to only some strongest subset of edges. Furthermore, this pattern holds for experiments with $75$, $150$, or $300$ samples (see supplemental figures).

\begin{figure}
    \centering
    \includegraphics[scale=0.25]{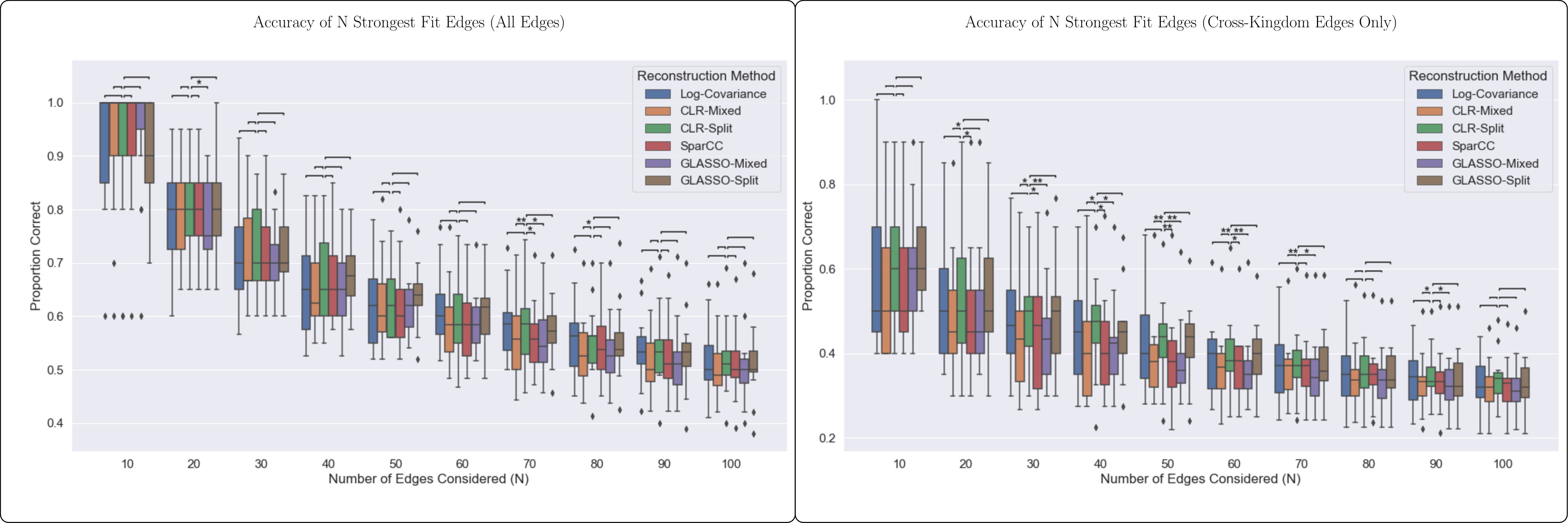}
    \caption{For experiments with $150$ samples, all six methods performed equally well in accuracy of their top strongest edges, using both all edges and cross-kingdom edges only. We present the significance of the difference between the ``CLR-Split" method and the other methods to emphasize the importance of addressing the error of cross-kingdom covariance estimation.}
    \label{fig:acc}
\end{figure}

The second metric aims to assess each network's utility in identifying interactions for potential future study. In this context, the most relevant metric should reflect the likelihood that an inferred edge with a strong magnitude is an edge in the ground-truth network used to generate the simulated data. To measure this, we computed the accuracy of the top $N$ strongest inferred edges for each method, with an edge being called ``correct" if it appears with the same sign (i.e. both positive or both negative) in the ground-truth network. \Cref{fig:acc} reveals that there is no significant difference between methods when it comes to this metric. This suggests that a researcher's confidence in the strongest edges of the fit networks does not depend on the network inference method that is used.

\subsection{Hub Node Detection}

We measured each method's ability to identify nodes as network hubs, which may be related to keystone taxa in the community. Keystone species are taxa that have a significant impact on the organization of a microbial community\cite{mills1993keystone,herren2018keystone}. Modern methods to identify keystone species make use of the data beyond reconstructed networks with techniques based on machine learning\cite{amit2023top} or inferred dynamical systems\cite{berry2014deciphering}. Due to our focus on network reconstruction, we investigated methods to detect network hubs as defined by the centrality of nodes. Centrality is a concept of how connected a node is within a network, and can be measured using a variety of metrics. Most common is ``betweenness centrality", which measures how often a shortest path through any two nodes passes through the node of interest\cite{hagberg2008exploring}. Hub nodes are often considered possible ``keystone" species\cite{banerjee2016determinants,vick2014modular}, although we note that they need not be. In addition to betweenness centrality, we used two other common metrics for centrality. These are ``degree centrality", which is simply based on the number of edges connected to a node, and ``eigenvector centrality", which is related to a random walk on the network\cite{hagberg2008exploring}. Finally, a reconstructed covariance network can be taken to be a set of parameters for the Lotka-Volterra dynamical system, although we note that this will be unlikely to match the parameters fit to time-course data. This allows us to compute a ``keystoneness" score based on the impact of removing a taxa from the community in a simulated knock-out experiment using Lotka-Volterra dynamics\cite{berry2014deciphering}.

\begin{figure}
    \centering
    \includegraphics[scale=0.22]{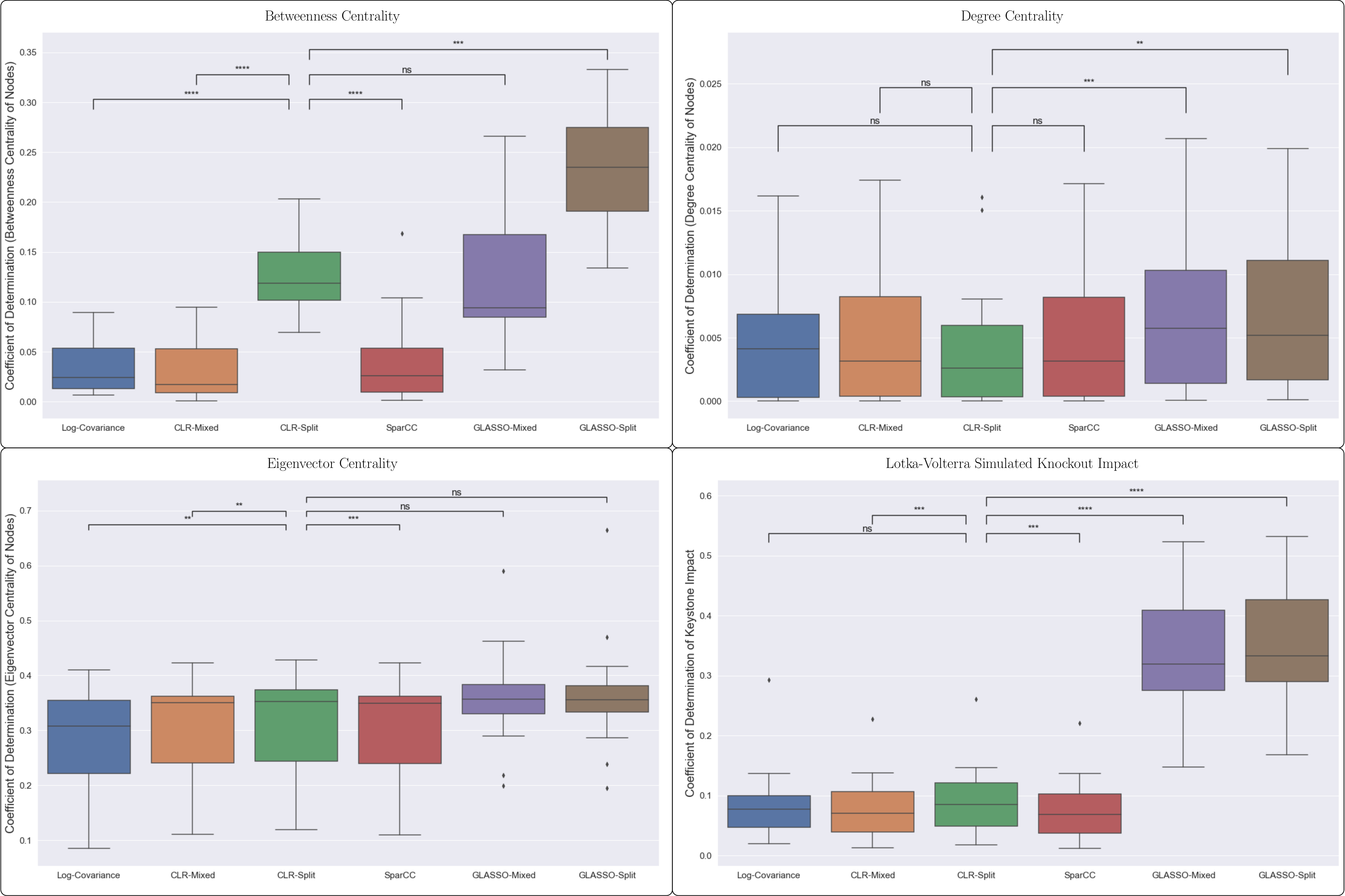}
    \caption{Hub/keystone species classification varied significantly based on the method used to define a hub or keystone species. Accounting for the error in cross-kingdom covariance made slight improvements to the agreement between ground truth and reconstructed hub scores when hubs were defined by betweenness centrality, degree centrality, eigenvector centrality, and a Lotka-Volterra dynamical system, but not when they were defined by degree centrality. However, using the more sophisticated ``GLASSO" methods improved hub classification for all methods.}
    \label{fig:hubauc}
\end{figure}

In \cref{fig:hubauc}, we compare the hub or keystone score for each node in the reconstructed network with its score in the ground-truth covariance network, with scores defined by betweenness centrality, degree centrality, eigenvector centrality, and simulated knock-out experiments. We show the coefficients of determination for this comparison across 20 trials for each reconstruction method.
 
\subsection{Performance of Community Detection}

\begin{figure}
    \centering
    \includegraphics[scale=0.22]{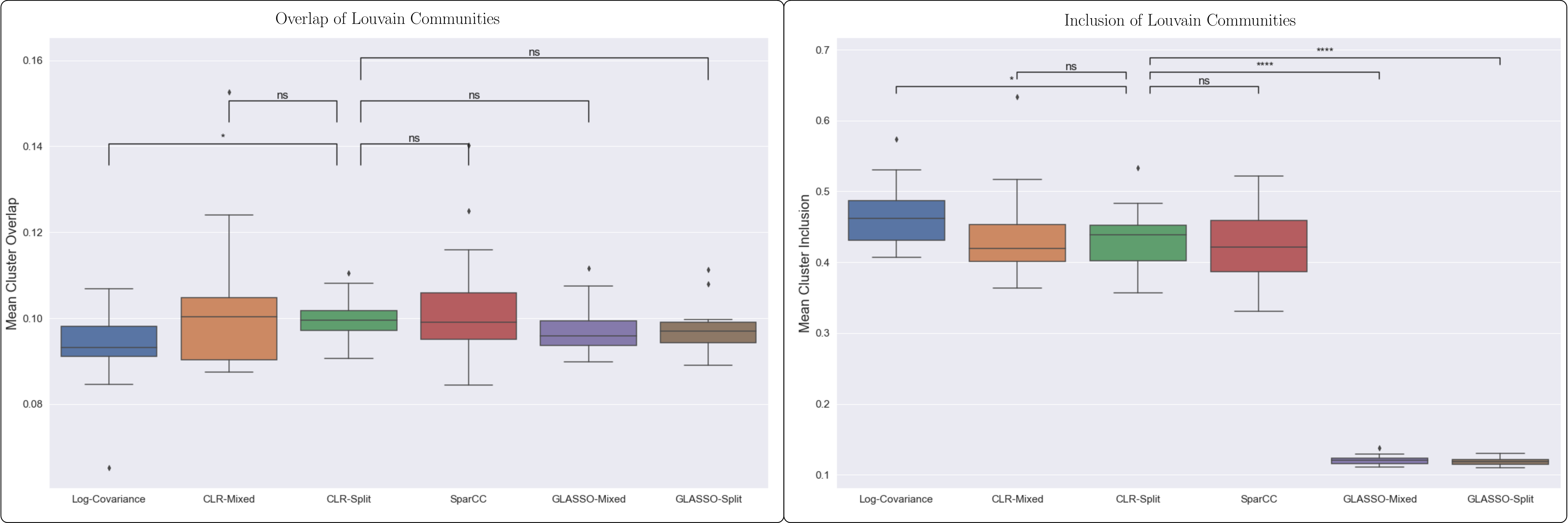}
    \caption{Reconstruction method did not significantly impact the detection of communities using network-based community detection when measured by the overlap of communities. When measured by community inclusion, defined as the extent to which a community from the reconstructed network was included in a single community in the ground truth network, the ``GLASSO" methods performed poorly in comparison to the other methods. In either case, accounting for the error in cross-kingdom covariance did not effect the results.}
    \label{fig:clusters}
\end{figure}

Reconstructed covariance networks are often used to define microbial communities\cite{jackson2018detection,shaffer2023scnic}, and so we tested each network reconstruction's community structure for agreement with the ground truth covariance network. We defined communities in each network using the Louvain community detection algorithm\cite{blondel2008fast}. We compared communities detected in the reconstructed networks with the communities in the ground truth network in two ways, both of which assigned scores to the communities in the reconstructed networks. First, we found for each community in the reconstructed network the community in the ground-truth network that it most overlapped with and scored it by the size of the intersection as a proportion of the size of the union of the two communities. Next, we found the community in the ground-truth network that contained the most nodes from the community in the reconstructed network and scored it by the size of the intersection as a proportion of the size of the reconstructed network community. \Cref{fig:clusters} shows that reconstruction method did not have a significant impact when community detection was evaluated according to overlap, while the ``GLASSO" methods performed poorly when community detection was evaluated according to inclusion.

\subsection{Performance Using Real Data}

\begin{figure}
    \centering
    \includegraphics[scale=0.2]{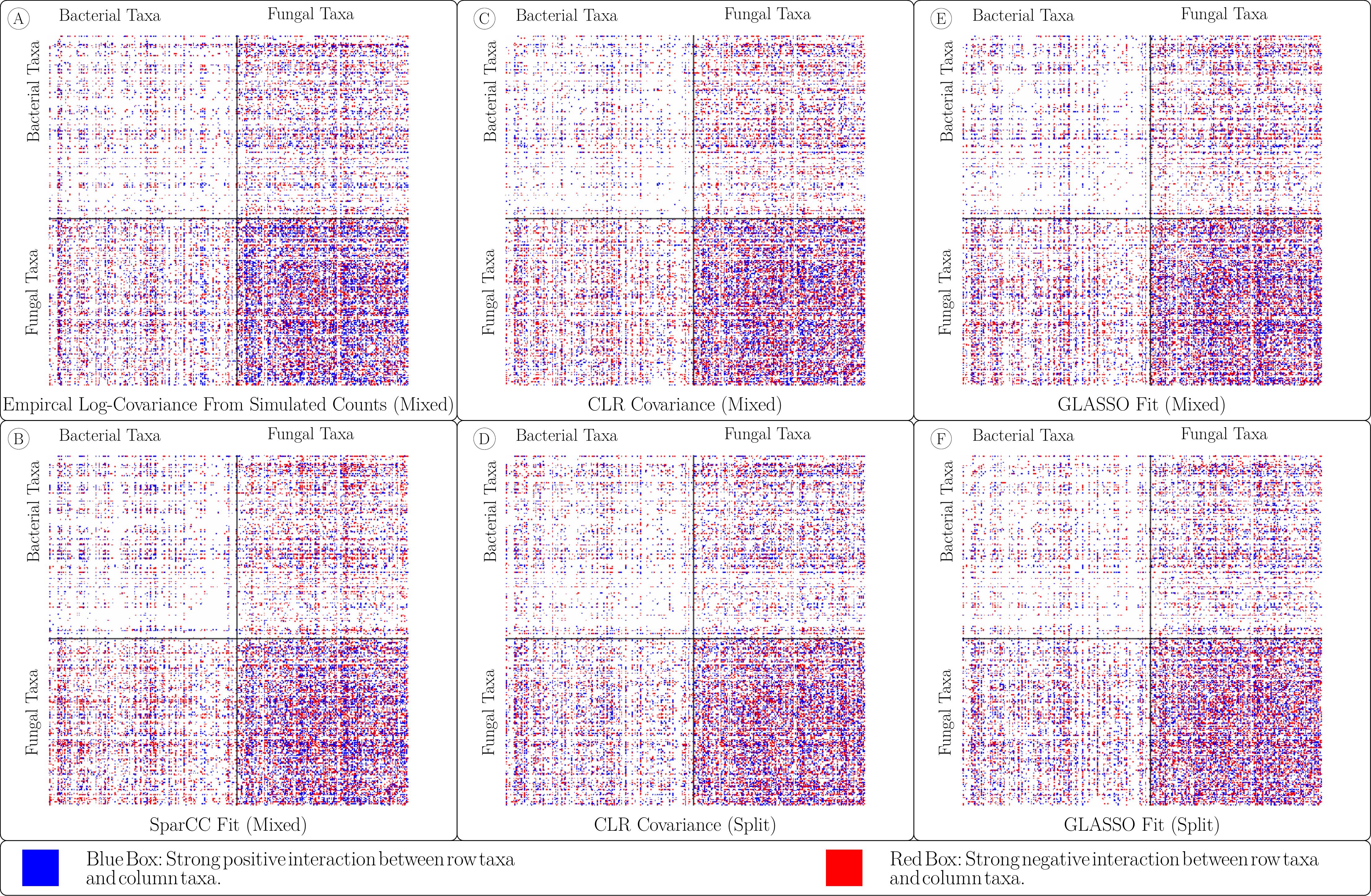}
    \caption{The networks generated using the six methods for reconstructing the covariance matrix for data from drought-effected soil mesocosms published in de Vries et al.\cite{de2018soil}. The top $10\%$ of edges by absolute strength are shown color coded as red for a negative interaction and blue for a positive interaction. While only networks (D) and (F) account for the error introduced by combining 16S and ITS data, the bias of the kind we see in \cref{fig:nets} and \cref{fig:meanedge} is weak.}
    \label{fig:realnets}
\end{figure}

The networks generated from the soil mesocosm data published in de Vries et al.\cite{de2018soil} do not show obvious bias. However, comparing mean edge strength that the pattern observed in synthetic data and shown in \cref{fig:meanedge} does hold to a small degree, except in the case of the log-covariance network. \Cref{fig:realnets} shows a representation of the covariance matrices estimated from that data. While it is not apparent to the naked eye, we observed a slight but statistically significant bias with the same appearance as the synthetic data. Mann-Whitney U test p-values comparing edge strength distribution of intra-kingdom and cross kingdom edges were $<0.05$ for the Log-Covariance method (fungi vs cross-kingdom only), CLR-Mixed method, SparCC method, and GLASSO-Mixed methods. The two likely reasons for this bias to be so slight: (1) a relatively constant biomass ratio between bacterial and fungal kingdoms and/or (2) other sources of error at high enough magnitude to obscure the error due to combining data types.

\section{Discussion}

It is clear from a relatively straightforward theoretical calculation (\cref{eq:xking_err}) that combining paired data sets without regard for their compositional nature introduces error when trying to infer a covariance network from the data. Furthermore, there exists a solution to this problem, which is to compute the centered log-ratio of the two data sets separately before combining them. In fact, the best performing methods in our analysis are the two that use this solution, the CLR-Split method that simply takes the covariance of the resulting transformed data, and the GLASSO-Split method that starts with that covariance and further refines it by a maximum-likelihood procedure. In light of the fact that the GLASSO method should in general improve our estimate by eliminating dependence among covariances (i.e. forcing the fit matrix to be full rank) and enforcing realistic sparsity constraints, it seems clear that this is the best method to use in constructing transkingdom networks from paired compositional data sets. 

However, the GLASSO method requires solving an optimization problem called a ``semi-definite program", which is significantly more computationally taxing than other methods of network fitting. Thus, if a different method can be used with acceptable results, it may be preferable to the GLASSO method for practical reasons of computational power and time. It is therefore worth considering the various sources of error in network fitting, and deciding if the error introduced by concatenating two data types from a paired data set will make a major impact on what is being inferred. Perhaps not surprisingly, this will depend on the network's intended purpose and the richness of the data set, as computational time scales with number of taxa for each method.

Obviously, we cannot quantify error exactly in networks reconstructed from real data, because we do not know the underlying truth. However, we can look for signature of the error of combining paired data sets by inspecting the difference in distribution between intra- and cross-kingdom edges. We observe in synthetic data that intra-kingdom edges tend to be biased in a positive direction when paired-data error is not accounted for, and to a lesser extent cross-kingdom edges tend to be biased in a negative direction. We observe in real data that this pattern is indeed followed. It is impossible to say with certainty that differences in intra- and cross-kingdom edge distribution are due to the errors introduced by combining data sets, but our analysis suggests that they may be.

Our analysis has focused only on reconstructing the covariance between the taxa in a set of samples. To understand more general relationships, including across kingdoms, other methods should be employed alongside covariance network reconstruction. In particular, non-negative tensor factorization methods show promise in identifying relationships between groups of taxa. Two popular tools, MOFA+\cite{argelaguet2020mofa} and DIABLO\cite{singh2019diablo} are both designed to use non-negative tensor factorization to identify groups of related variables across data types. 

We note that two experimental methods do exist which might eliminate the problem of estimating covariance across kingdoms by providing correctly normalized data. The first is to use metagenomic sequencing for taxonomy\cite{freitas2015accurate,blanco2023extending,wood2019improved}. This technology classifies taxa by comparing reads to a known database in order to assemble genomes for the taxa present in the community\cite{quince2017shotgun}. Because metagenomic sequencing does not target a specific gene, it can in theory provide an estimate of relative abundance across kingdoms. Furthermore, metagenomic sequencing has been demonstrated to improve taxonomic identification of bacteria when compared to 16S amplicon sequencing\cite{odom2023metagenomic}. In this case, the problem of inferring covariance networks is reduced to the single data-set case. However, practical problems with using shotgun metagenomic data to identify the relative abundance of fungi and bacteria must be addressed. For instance, because shotgun metagenomic sequencing does not make use of a true ``bar code" region, closely related organisms may be grouped together\cite{liu2022opportunities}, and results can be misleading if sequence abundance is confused with taxonomic abundance\cite{sun2021challenges}. Additionally, because metagenomic sequencing does not include an amplification step, rare taxa can be easily missed. Finally, most metagenomic sequencing analysis platforms are optimized for bacterial taxonomy identification, leading to error taxonomy identification across for fungi\cite{usyk2023comprehensive}, or possible bias when separate methods are used for fungi (e.g. FindFungi\cite{donovan2018identification}).

The other experimental method that could allow combining data sets is to measure absolute abundance of the taxa in the sample. In this case, all of the problems of using compositional data are eliminated. Recent advances in estimating absolute abundance, especially of bacteria, have been achieved by a variety of methods including qPCR\cite{bonk2018pcr}, flow cytometry\cite{props2017absolute}, and the use of an ``internal standard"\cite{zemb2020absolute}. However, significant drawbacks exist with all of these, including cost, difficulty, and a lack of accuracy\cite{harrison2021quest}. To further complicate the problem, fungal biomass is generally estimated using a biomarker, ergosterol (5,7-diene oxysterol)\cite{adamczyk2023step}. This method involves the use of liquid chromatography, adding a technical hurdle and expense, and can vary in effectiveness depending on extraction method\cite{wilkes2023ergosterol}. Additionally, using separate methods to infer absolute abundance for two different kingdoms will likely produce unpredictable biases when the data from the two kingdoms is combined.

If we assume that the underlying true distribution of the taxa in a set of samples is indeed log-normal and attempt to infer the corresponding covariance matrix, and so network of interactions, we can characterise the error we introduce in network inference. In that (already ideal) case, regular network inference (i.e. from a single compositional data-set) introduces three sources of error:
\begin{enumerate}
    \item we use a finite number of samples, so estimating the underlying distribution from the empirical covariance of absolute log-abundances is only approximate,
    \item our data is compositional and not absolute, so estimating the empirical covariance of absolute log-abundances is only approximate,
    \item we use a finite number of reads for each sample, so estimating the relative abundances for each sample is only approximate.
\end{enumerate}
Using paired compositional data introduces a fourth source of error, which we can eliminate using, for example the GLASSO-Split method. However, if this last source of error is small enough compared to the first three, it may be unnecessary to spend the computational resources to eliminate it.

In fact, if our goal is to identify possible interactions for further study, any of the network methods that we tested appear to be adequate. Furthermore, if we are looking for communities in our network, each method performs roughly equally, while accounting for paired-data error provides some improvement in identifying important taxa as hub nodes. Which method should we use, then? In considering an answer to this question, we should consider the fact that the SparCC and GLASSO methods enforce realistic sparsity constraints and eliminate linear dependence among in the inferred covariance matrix, and so may perform better on real data than is suggested by our experiments. We should also note that if we wish to identify edges, we need to choose some cutoff in strength in the inferred covariance matrix that we will call an existing edge. This can be done by a bootstrapping approach that provides an estimated probability that an edge above a threshold strength might occur randomly (a p-value)\cite{faust2012microbial}. Such a bootstrapping procedure requires repeated network inference on randomized data, and so multiplies the computational cost of network inference.

If time and compute power allows, or the number of taxa considered is small, the GLASSO-Split method likely provides the best estimates of true transkingdom covariance networks from paired compositional data of the methods that we considered. However, it also carries the highest computational cost. The error introduced by concatenating paired data sets depends on the ratio of the total biomasses of each kingdoms being measured in each data set (e.g. ratio of total bacterial biomass to total fungal biomass), or more precisely the variance of this ratio and the covariance of this ratio with taxa abundances. If this quantity can be estimated or reasonably assumed to be close to constant, then there is no added error in concatenating data sets without correction. Finally, our analysis suggests that simply computing the covariance of data that has been separately CLR-transformed and then concatenated provides a reasonable starting point to look for interactions and keystone taxa.

\section*{Acknowledgements}

This research was supported by a Science Focus Area Grant from the U.S. Department of Energy (DOE), Biological and Environmental Research (BER), Biological System Science Division (BSSD) under grant number LANLF59T.

\section*{Competing Interests}

The authors declare no competing interests in relation to the work described.

\section*{Availability of Code \& Data}

Our code to generate synthetic data and assess networks, as well as the synthetic data generated for our analysis and the networks generated from that data, are available through the EDGE Bioinformatics platform and can be found at \url{https://github.com/LANL-Bioinformatics/synthetic_transkingdom_data}. Our GLASSO methods were computed using the Julia language's JuMP package\cite{Lubin2023}. Code for this implementation is available at \url{https://github.com/LANL-Bioinformatics/normalCowboy}

\end{document}